# From Stacks to Circuits: A Regenerative Socio-Technical Roadmap for AI Infrastructure within Planetary Boundaries

This document is a working paper and reflects the state of research as of May 2026. Comments are welcome and should be directed to the corresponding author at h.liao@ieee.org.

Han-Teng Liao*
*Independent Researcher*

Penang, Malaysia
h.liao@ieee.org
0000-0003-1081-5599

Karen Ang
*Independent Researcher*

Kulim, Malaysia
karenang118@gmail.com
0009-0008-5923-0106

***Abstract*—** Current scaling trajectories for Generative AI, typified by linear supply-side "stacks," prioritize performance density while externalizing significant thermodynamic and material costs. As the "Twin Transition" of green and digital transformation accelerates, the industry faces technology gaps—including Scope 3 emissions and e-waste recycling—that impede sustainable scaling and lead to social tensions. This study proposes a Regenerative Socio-Technical roadmap that repurposes the Sustainable Production and Consumption system map to reframe artificial intelligence infrastructure as a system-of-systems governed ultimately by planetary limits. By integrating the Institute of Electrical and Electronics Engineers International Roadmap for Devices and Systems (IEEE IRDS) sustainability considerations for semiconductor facilities, the study proposes a metabolic circuit framework that centers "Values and Needs" within production and consumption relationship loops. This study identifies critical gaps in current Nvidia-centric roadmaps and proposes a competing reference architecture. It demonstrates how a spontaneous order of resource parsimony and planetary accountability can provide an actionable pathway for regulatory compliance and industrial resilience in the digital circular economy.

***Keywords*— *Carbon emissions; Environmental management; Green manufacturing; Circular Economy; Roadmapping; Socio-Technical Systems; Scalability; Twin Transition***

## I. INTRODUCTION: FROM STACKS TO CIRCUITS

### *A. Context: Rapid and Intensive Scaling of Generative AI*

The rapid scaling of Generative Artificial Intelligence (AI) infrastructure, typified by linear "stack" models such as the Nvidia "five-layer cake" architecture [1], [2], [3], has outpaced the development of governance frameworks capable of managing its systemic environmental footprint. This intensive scaling, which treats the AI stack as an optimized pipeline from "Electrons to Tokens", is driven by a performance-at-all-costs paradigm that prioritizes responsive fluency over thermodynamic efficiency and environmental sustainability.

This scaling trajectory imposes significant stress on the existing technology gaps identified in the 2024 Institute of Electrical and Electronics Engineers (IEEE) International Roadmap for Devices and Systems (IRDS™) regarding the Environment, safety, health & sustainability (ESHS) of the Environmental Sustainability of the Semiconductor Facilities (ESSF) [4]. Specifically, the roadmap emphasizes that facility-level resource consumption must consider the interconnectivity of water and energy, and the roadmap itself represents the collaborative outcomes by expert representatives of leading semiconductor manufacturers, original equipment manufacturers (OEMs), materials suppliers, analytical labs, and consultants. The roadmap also specifies quantitative targets in order to support "building climate-resilient facilities, meeting stringent emission limits, and managing large and growing footprints are all critical objectives the industry faces" (p.4)

Empirical data, particularly manufacturing and environmental impact metrics, provide science-based evidence for systemic evaluation. For example, the development of Digital Product Passports (DPPs) aims to assist industrial managers in accounting for operational decision-making within reverse supply chains [5]. DPPs are widely expected to serve as core enablers of the circular economy [6], demonstrated by recent regulatory models in the EU Battery Regulation [7] and the textile industry [8]. Consequently, the European Commission regards the DPP as a fundamental mechanism for driving this transition towards a product level circular economy [9], [10], [6], [11].

As Europe actively pursues the "Twin Transition" of green and digital transformation, the digital circular economy (DCE) has emerged as a critical nexus [6], [7] for ensuring that technological growth remains within planetary boundaries [12]. Thus, the integrated demands of ESHS issues require rigorous alignment between digital innovation and material circularity. Without this synchronization, any Twin Transition initiative will risk becoming a ecological liability where efficiency gains accelerate resource depletion. This alignment is especially critical at a time when the global economy is driven by capital expenditure (CapEx) on generative AI infrastructure that is inherently energy- and resource-intensive.

### *B. Motivation: What if Generative AI is Not Regenerative?*

A fundamental misalignment persists between the financial valuation of AI as an infinite digital asset and the physical reality of its resource-bound production and consumption patterns. This study argues that the current human-machine metabolism prioritizes computational throughput at the expense of planetary stability and cohesive social relations. While the prevailing industry narrative, championed by figures such as Nvidia CEO Jensen Huang, frames the AI stack as an optimized journey from "electrons to tokens" [1], [2], [3], this paper interrogates the long-term viability of high-throughput tokenomics through the lens of an "atoms-to-values" framework grounded in material circular circularity. Tokenomics needs such physical ground truths to remain sustainable.

A primary driver of this misalignment is the Jevons Paradox: as energy and hardware efficiency gains reduce the unit cost of tokens and compute, aggregate demand expands at a velocity that outpaces those efficiency gains, resulting in a net increase in total energy and material consumption. This rebound dynamic is clearly observable at the regional infrastructure level. In Taiwan, Jensen Huang asserted that "human labor needs rice, but AI labor needs electricity" [13] , calling for expanded energy capacity in Taiwan in May 2026. It is coincided with state utility Taipower enforcing strict regulatory ceilings on new data center grid applications north of Taoyuan, due to acute supply-demand imbalances between southern generation capacity and northern demand [14]. This Grid Paradox illustrates how linear scaling assumptions collide with physical and regulatory constraints. It underscores the urgency of moving beyond static corporate social responsibility (CSR) reporting toward dynamic system models that treat waste heat, water, and embodied carbon as recoverable liabilities rather than sunk costs.

The social dimensions of these socio-economic tensions have attracted equally high-level institutional scrutiny. Pope Leo XIV's 2026 Encyclical, *Magnifica Humanitas* [15], explicitly identifies underpaid data labelling, content moderation, and rare-earth and critical mineral extraction as "new forms of slavery," highlighting the human and environmental cost underpinning AI scaling. These structural issues force a critical question: What if generative AI is not regenerative?

Addressing both the thermodynamic and socio-economic dimensions requires embedding reflective governance and regulatory technology (RegTech, the use of technology to automate and deliver regulatory compliance) directly within the AI stack. Challenging "efficiency as ultimate value" paradigm, this architecture must align with the EU Corporate Sustainability Due Diligence Directive (CSDDD) and the Safe and Sustainable by Design (SSbD) guidelines [16]. EU CSDDD requires companies to conduct human rights and environmental due diligence across value chains, whereas SSbD advances frameworks for eliminating hazardous substances and processes.

This alignment ensures that both the "AI *of* the environment" (its physical and environmental footprints) and "AI *for* the environment" (its application toward environment and sustainability objectives) [17] are rigorously designed, monitored and audited. Ultimately, tokenomics must be re-engineered to support Industry 5.0 practices that balance the Triple Bottom Line (TBL) of profit, people and the planet.

### C. Objectives: Sensible, Technical, yet Holistic Roadmap

This study synthesizes regenerative accounting and socio-technical systems (STS) theory into a unified framework for computing economics and infrastructure governance.

The central research question is formulated as follows: *How can an integrated Regenerative Socio-Technical (RST) framework enable AI infrastructure governance to remain within planetary and regional resource boundaries while meeting regulatory mandates, such as CBAM and SSbD?*

By formalizing ecological debt as an observable balance-sheet liability and systemic resilience as a trackable engineering parameter, this roadmap introduces a definitive reference architecture for transitioning toward a sustainable, regenerative digital circular economy.

## II. EXISTING THEORIES & PREVIOUS WORK

This section bridges the gap between industry execution —represented by the five-layer stack— and strategic roadmapping, addressing the necessity of a rigorous sustainability audit. This includes using integrating IRDS parameters and, feedbacks and value-alignment within Sustainable Production and Consumption (SPaC) loops [18].

### A. Electrons to Tokens: Scaling Through Orchestration

Any new conceptualization of a DCE must contend with Nvidia's scaling-centric strategy, given its primary role of the AI value chain in 2025-2026. Nvidia's five-layer cake (Energy, Chips, Infrastructure, Models, and Applications) is a high-throughput pipeline designed to commoditize energy and maximize the value of a token [1], [2], [3]**.** This mental model treats the data center as a factory where electricity is the raw material and the final product is the token —the fundamental unit of AI intelligence. The industry thus shifts from traditional "atoms to values" manufacturing toward a high-throughput "electrons to tokens" tokenomics model. Nvidia's five-layer cake functions as a linear scaling framework that treats computing as an extractive process, assuming that efficiency gains justify increased total throughput.

### B. Critique: Sustainability and the Planetary Limits

Spurred by AI workloads, data center electricity consumption is projected to double by 2026, reaching total energy demand level of Germany [19]. This trajectory contributes to the global e-waste stream [20], [21], [22], even as AI is proposed as a tool to manage that waste [23], [24]. The "electrons-to-tokens" paradigm remains socio-technically blind, as shown in TABLE I. By treating energy as a static utility and lacking circularity, it ignores the regenerative relations needed to sustain the human and material base.

TABLE I. THE 5-LAYER AI PRODUCTION STACK (ADAPTED FROM [1])

| Layer | Functional Domain | Industrial Role |
|---|---|---|
| L5 | Applications | End-user services (e.g., Enterprise Agents, Robotics, and Smart Apps) |
| L4 | Models | Cognitive engines (e.g., LLM) |
| L3 | Infrastructure | AI and Data Centers (e.g. compute) |
| L2 | Chips | Semiconductor hardware (GPU/TPU/NPU acceleration) |
| L1 | Energy | Power generation and cooling: often thermodynamic and resource input |

Consequently, this electrons-to-tokens linear view rests on resource assumptions that face structural friction at the Energy, Chips, and Infrastructure layers. These standards and critiques preceded Large Language Models (LLMs). Under Recommendation ITU-T L.1480 ("Enabling the net zero transition: Assessing how the use of ICT solutions impacts greenhouse gas emissions of other sectors") [25]. this narrow scaling view fails systemic auditing, especially regarding higher-order impacts such as rebound effects, via the Jevons Paradox. Viewed through a System-of-Systems (SoS) lens [26], as token unit economics decline, an AI Agent boom can trigger non-linear shifts in aggregate demand. This rebound effect expands macro-energy consumption beyond the structural adaptation velocity of local power networks, posing severe risks to socio-environmental sustainability.

### C. Baselines, Grid Volatility, and Energy Communities

AI demand escalation challenges utility supplies. Rapid AI and data center investments reshape global load profiles, forcing energy security, affordability, and sustainability boundaries to shift rapidly [27]. Modern hyperscale data centers exhibit highly volatile electricity and water use, intensifying frictions between digital growth and sustainable development [28]. Aggressive procurement of clean energy to claim operational carbon neutrality on paper pushes energy policies toward "all-of-the-above" options, including nuclear power, while externalizing risks and costs onto local grids [29].

To mitigate these disruptions, power systems research highlights Energy Communities as vital for local grid resilience and decarbonization [30], [31], [32]. Transitioning from passive consumers to proactive prosumers with localized generation and storage, these communities allow active end-users to share energy virtually within a defined geographic perimeter. Their viability depends on enabling policy schemes, practitioner support frameworks, and coordinated aggregator business models.

### D. Regenerative Accounting: Managing Ecological Debt for Systemic Reliability within Sustainability Limits

The current state of AI adoption is characterized by upstream and downstream ecological debt [33], accelerating across supply chains through acute grid stress, rapid e-waste accumulation, and mounting thermal entropy [34], [35]. Green AI narratives focused on algorithmic optimization and offsets can obscure the material reality: hyperscale data centers treat biophysical systems as extractive pools, drawing massive water volumes for cooling and accelerating the obsolescence of advanced hardware. The linear "electrons-to-tokens" model operates detached from the finite carrying capacities of ecological systems, creating metabolic friction that undermines long-term resilience. The UNU Global E-waste Monitor highlights that e-waste is the fastest-growing waste stream globally [20], a trend exacerbated by the short replacement cycles of GPUs and components for AI centers.

This study builds on Regenerative Accounting, adapted here into a closed-loop biophysical optimization framework for digital infrastructure [36]. Rather than retroactively logging resource depletion as a business cost, a regenerative model embeds ecological restoration obligations and feedback loops as active functions in system accounting [37]. Nature is treated as an institutional actor whose regenerative capacity must govern operations. Just as generative models are deployed to assess risks in debt-for-nature swaps [38], a reciprocal computational logic is applied inward to the computing stack to audit token production against regional infrastructure limits.

By synthesizing these accounting pillars with the material turn in ecological macroeconomics, this study proposes an enforceable eco-debt ceiling. This construct codifies lifecycle ecological liabilities—such as mineral extraction, water depletion, and digital waste—into system-level constraints [35], [37]. Enforcing a ceiling on compute throughput based on real-time environmental baselines injects sufficiency, accountability, and reparative redistribution into the orchestration layer. The architectural layers are thus compelled to operate within localized constraints that protect civil resource sovereignty, respect planetary limitations and implement circular economy policies.

Furthermore, the manufacturing phase of advanced semiconductors consumes large volumes of water and generates persistent chemical waste, creating a metabolic lock-in that layered linear stacks cannot address [39].

### E. The IEEE IRDS™ lens: Sustainability Technology Gaps

Sustainability-oriented semiconductor manufacturing requires benchmarking against planetary limits [40]. As the successor to the International Technology Roadmap for Semiconductors (ITRS), the IEEE International Roadmap for Devices and Systems (IRDS™) aggregates expert predictions for the next fifteen years of electronics development. The 2024 IRDS edition includes an updated Environment, safety, health & sustainability (ESH/S): Environmental sustainability of semiconductor facilities (ESSF) roadmap chapter [4].

Foundational work like this shifts industry from reactive compliance toward proactive, design-centric principles consistent with SSbD. Three key pillars are relevant to this study:

- IRDS ESH/S ESSF Roadmap [4]: Specifies Key Performance Indicators (KPIs) for water and energy efficiency and promotes the minimization of hazardous chemistries at the process design phase, including Green Tool Characteristics.
- Framework for Critical Semiconductor Materials [41]**:** Develops requirements to assess critical materials (e.g., advanced lithography) based on toxicity, circularity, and eco-impact before high-volume deployment.
- Novel Sensing for Environmental Solutions [42]: Extends SSbD to applications by designing devices with high selectivity and low power for real-time resource monitoring.

For engineering managers and technology strategists, the 2024 IRDS ESSF signals that sustainability is now a mission-critical enabler for growth. AI-driven demand for larger, more complex facilities intersects with resource scarcity (water and energy), tightening regulations (e.g., per- and polyfluoroalkyl substances [PFAS], greenhouse gas [GHG] limits), and physical climate risks [43], [44]. The widening technology gaps identified in the roadmaps will be stressed by surging global demand in AI, electric vehicles (EVs), and high-performance computing, forcing managers to balance economic scalability, supply chain resilience, and environmental sustainability [45].

### F. The Sustainability-Scaling Gap

While Nvidia's five-layer cake remains a dominant blueprint, it operates as a linear, extractive model decoupled from the physical and regulatory boundaries. A critical gap persists: the industry lacks a unified "Sustainability Audit" to reconcile the resource intensity of tokenomics with constraints such as the Carbon Border Adjustment Mechanism (CBAM) and SSbD regulations. CBAM prices the carbon content of imported goods to prevent carbon leakage [46]. Without a framework linking downstream demand to upstream resource limits, the industry risks hitting a hard ceiling that halts industrial growth due to environmental non-compliance or resource exhaustion.

This study addresses this gap by introducing an integrated Regenerative Socio-Technical (RST) framework. This architecture functions as a bidirectional metabolic loop, translating static IRDS environmental parameters into live operational telemetry within SPaC relationship loops. By grounding computational demands in regional grid limits and regulatory thresholds, the RST framework shifts governance away from unconstrained scaling toward verifiable, bounded utility management, directly answering the socio-technical and regulatory objectives of this study's roadmap.

## III. METHODOLOGY

### A. *Presented Study, Research Question(s), and Hypothesis*

This study investigates the systemic tension between the critical planetary boundaries and resource-intensive, extractive nature of generative AI infrastructure. It addresses the core inquiries formulated at the end of Section II, seeking to establish an integrated auditing framework that reconciles linear "tokenomics" with environmental constraints.

The research is guided by the hypothesis that a Regenerative Socio-Technical (RST) framework—which integrates environmental factors (e.g., IRDS parameters) into the core of Sustainable Production and Consumption (SPaC) relationship loops [18]—can effectively mitigate Scope 3 emissions and fulfill SSbD mandates while sustaining digital infrastructure growth for human development.

### B. *Repurposing SPaC loops as RST framework*

To address the systemic tensions between generative AI infrastructure and planetary boundaries, this study repurposes the traditional SPaC framework into an RST system dynamics model. To better conceptualize modern digital infrastructure driven by AI compute, the proposed model introduces sector-specific node definitions tailored to deep-tech hardware, establishes a clear specification of environmental mandates, and embeds explicit system dynamics feedback structures.

Fig. 1, the proposed "Atom-to-Values" SoS model maps the system dynamics of compute a TBL taxonomy. It aims to balance profit, people and the planet for sustainable growth, thereby providing a circular reference:

**Profit (The Core Capital Loop):** The inner cycle tracks the economic engine of AI hardware.

- *Extraction* represents the input of critical minerals, water, and energy;
- *Production* corresponds to semiconductor fabrication and AI generation;
- *Distribution* covers hardware logistics, digital networks, and software platforms;
- *Consumption* encompasses smart apps and consumer agents; and
- *Investment* reflects massive capital expenditure (CapEx) reinjected into compute capacity building.

At the epicenter, the Values/Needs (Demand) of chips, compute, and hyperscale systems act as a socio-technical governance anchor where value-based requirements—such as KPIs and SSbD criteria—shape demand and funding priorities.

**Planet (The Ecological Ceiling and Recovery):** Centered on the *Extraction* and *Waste (impacts)* nodes, this dimension enforces two distinct environmental system dynamics that enforce physical boundaries: *(a) Oscillating Dynamics:* A critical feedback loop characterized by material and information delays, capturing the systemic risks of ecological disruption. *(b) Recycling Dynamics:* A closed-loop material recovery pathway engineered to mitigate ecological strain and risks.

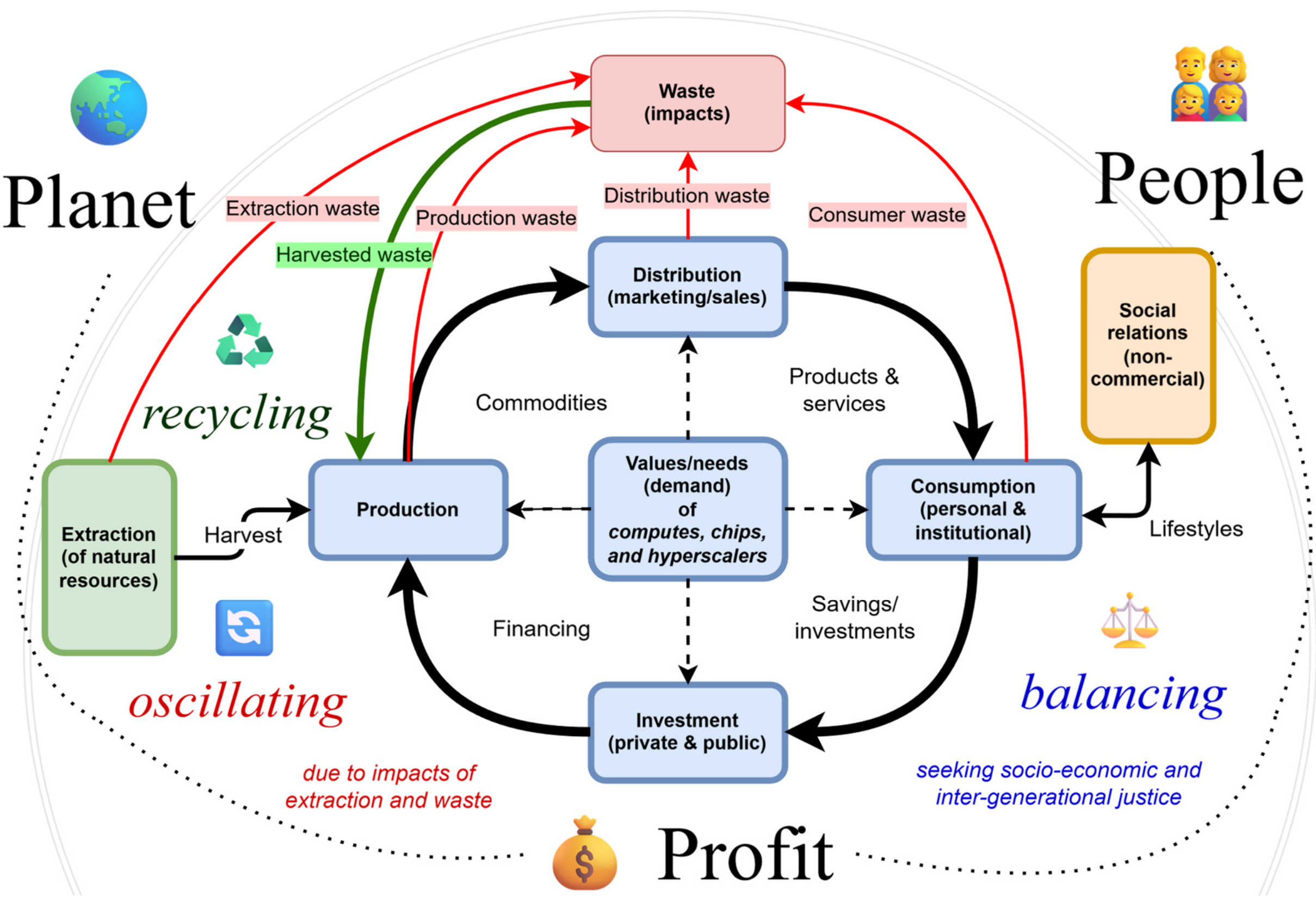

Fig. 1. The "Atom-to-Values" System of Systems (SoS) loops for AI compute (adapted and expanded from the Sustainable Production and Consumption (SPaC) framework by J. Barber in [18].

**People** (**The Socio-Technical Transitions**): Intersecting the nodes of *Social Relations*, *Consumption*, and *Investment*, this dimension captures societal governance through the last of the three system dynamics: (c) *Balancing Dynamics*. By aligning profit with socio-economic and inter-generational public procurement justice, it actively shapes lifestyle, saving, and investment behaviors to sustain healthy social and economic relationships.

### *C. Applied Research Methods*

By applying design science research principles—a methodology widely established in information systems research [47], [48], [49], this article generates a conceptual DCE framework. This framework details how the semiconductor and AI sectors produce digital deep-tech solutions for sustainable supply chains, where "digital deep tech" denotes advanced technical applications integrated with significant scientific or engineering innovations. This study views that, in high-tech manufacturing, green digital deep tech serves as the nerves of information, communication, and control, incorporating sustainable production, resource efficiency, and circular business models.

### *D. Data Collection*

The data collection protocol relies on secondary document analysis, ecosystem modeling, and industry standard mapping. The collected objects and artifacts consist of:

- **Industry Roadmap Data:** Technical specifications and energy consumption guidelines derived from the IRDS ESH/S ESSF roadmaps [4], which are expected to be shaped by CBAM and SSbD regulatory demands.
- **Tokenomics and Infrastructure Scaling Data:** Architecture requirements, performance metrics, and production layers surrounding the linear model.
- **Analytical Formulas:** Mathematical equations, performance curves, and operational scaling models for compute inference efficiency compiled from baseline literature [50], [51], [52], establishing quantified expressions of systemic impacts.

### *E. Analysis Method*

A theory-generating case study approach [53], [54] is applied by categorizing and resensitizing the linear structures of the semiconductor and AI sectors into accessible, circular system dynamics design reference architecture. Quantitative formulas of compute inference costs are strategically integrated to enrich the "Atom-to-Values" SoS loops, enabling a balanced qualitative and quantitative description of sustainable compute capacity.

This perspective integrates system dynamic insights from the architectural Roofline Model of compute throughput as a baseline variable for measuring thermodynamic efficiency and broader environmental impacts. By exploring how these analytical formulas can be integrated with Regenerative Accounting, this paper transforms sustainability requirements into drivers of generative innovation rather than mere compliance burdens. Shifting the design focus from raw compute density to "values-to-atoms" circular sensitivities formalizes ecological debt as an observable balance-sheet item and resilience as a trackable technical parameter, offering a definitive reference architecture for the transition toward a regenerative digital circular economy.

## IV. Results: Regenerative Socio-Technical Reference Architecture

By redefining the generative AI infrastructure stack as a metabolic loop, this study proposes an integrated an integrated Regenerative Socio-Technical (RST) framework.

### *A. Boundary Inputs for Metabolic Governance*

The first primary result of this study is the formal translation of multi-scalar telemetry data into an integrated baseline for automated system regulation. Rather than treating physical infrastructure limits, micro-architectural telemetry, and international policy mandates as disjointed operational layers, the RST architecture synthesizes them into a unified, real-time boundary matrix. This matrix establishes the foundational parameters for metabolic governance, defining the biophysical constraints under which the computing fabric must achieve operational closure. Table II illustrates these structural frictions as they exist within the 2026 infrastructure ecosystem.

Table II. Three-Tiered System Constraints (Baseline for Metabolic Governance)

| Layer | Primary Data Source | Operational Core Parameters | Homeostatic Threshold Examples |
|---|---|---|---|
| Physical Infra-structure | IEEE IRDS ESSF Roadmaps [4] | Water intensity, chemical toxicity, facility energy density, etc. | <0.5 liters per kWh compute; PFAS and emission controls [4] |
| Operational Compute | Inference Efficiency [50], [51], [52] | Model FLOP Utilization (MFU), memory wall bounds, High-Bandwidth Memory (HBM) capacity | Exponential thermodynamic efficiency drops at low concurrency regimes ($b(t) < b_{\min}$); $M_{\mathrm{KV}}$ memory sprawl |
| Policy & Governance | EU SSbD / CBAM [16], [46]; Papal Encyclical [15] | Border carbon adjustments, regional grid caps, labor exploitation audits. | Strict 5 MW grid allocation ceiling north of Taoyuan [14]; zero-exploitation supply chains [15]. |

To demonstrate the structural utility of these parameters, the RST reference architecture is explicitly contrasted against the prevailing linear "stack" model, which focuses narrowly on unidirectional electrons-to-tokens conversion [1], [2], [3]. Table III summarizes how such model overlooks critical externalized liabilities, which the RST architecture is engineered to internalize and regulate.

Table III. The 5-Layer AI Production Stack Liabilities

| Layer | Functional Domain | Key Externalized Liability (Systemic Metabolic Friction) |
|---|---|---|
| L5 | Applications | Uncapped token demand generation driving the Jevons Paradox rebound effect [25] |
| L4 | Models | Extreme density causing unmitigated HBM capacity sprawl [50], [51], [52] |
| L3 | Infrastructure | Cooling water depletion and unbuffered regional grid volatility [28],[29] |
| L2 | Chips | Embodied carbon of advanced lithography and e-waste lifecycles [20], [21], [22] |
| L1 | Energy | Extractive reliance on municipal grids, delaying green digital transitions [14], [29] |

Instead of treating energy and natural resources as infinite, exogenous supply inputs, the RST framework accounts for localized socio-technical and thermodynamic boundaries.

## B. *Compute Baselines and Thermodynamics*

The RST reference architecture recognizes that token generation within frontier AI accelerators is bottlenecked by a memory-to-compute imbalance (the "memory wall"). Shuttling model weights from HBM into local registers during sequential autoregressive inference creates a state of thermodynamic structural coupling, where physical silicon layouts dictate an application's metabolic energy profile. To amortize the thermodynamic cost of data transit, the system must maintain a minimum operational batch threshold ($b_{\min}$):

$$b_{\min} \approx K_{\text{hw}} \cdot \frac{P_{\text{total}}}{P_{\text{active}}}$$

where $K_{\text{hw}}$ is a dimensionless hardware constant expressing the ratio of peak performance (FLOPs) to memory bandwidth (bytes/sec)—empirically $\approx 300$ across 2026 accelerators—$P_{\text{active}}$ represents active layer parameters, and $P_{\text{total}}$ tracks the total parameter footprint. Falling below this threshold ($b(t) < b_{\min}$) drops the architecture into a low-concurrency regime that draws high static baseline power while idling on memory-bus transfers, generating high metabolic entropy per token.

This bandwidth bottleneck is compounded by Key-Value (KV) cache memory consumption ($M_{\text{KV}}$), which scales linearly with context length ($L$) as a biophysical liability:

$$M_{\text{KV}} = 2 \cdot b \cdot L \cdot E \cdot N \cdot \beta$$

where $b$ is batch size, $E$ is the hidden dimension, $N$ is network layer depth, and $\beta$ is precision byte density. At extreme frontiers, $M_{\text{KV}}$ surpasses the model weight footprint, forcing the orchestration layer to shard workloads across multi-GPU clusters to aggregate high-speed HBM capacity.

## C. *Systemic Volatility Indicator (SVI) Framework*

Unregulated, these two hardware constraints ($b_{\min}$ and $M_{\text{KV}}$) create a profound structural contradiction. Maximizing computational efficiency via optimal batch sizes ($b \geq b_{\min}$) causes the concurrent processing of extended context windows ($L$) to expand $M_{\text{KV}}$ linearly, triggering hardware out-of-memory faults. Conversely, down-batching ($b \to 1$) to preserve memory footprint traps the hardware in a memory-bound idling state, expending massive baseline power for sub-optimal throughput.

To resolve this tension and connect intra-chip constraints with regional energy networks, the architecture implements an automated metabolic governor mediated by a dimensionless Systemic Volatility Indicator (SVI), bounded at zero to prevent negative indexing:

$$\text{SVI}(t) = \max\left(0{,}1 - \frac{b(t)}{b_{\min}}\right) \cdot \Phi_{\text{grid}}(t)$$

where $b(t)$ is the instantaneous operating batch size, $b_{\min}$ is the hardware roofline threshold, and $\Phi_{\text{grid}}(t) \in [0{,}1]$ represents the localized power grid utilization ratio relative to institutional ceilings (such as Taipower's 5 MW regional allocation caps [14]). This indicator mathematically formalizes two primary operational states:

- **State A (Maximum Volatility):** When under-batched workloads ($b(t) \ll b_{\min}$) run during peak regional grid stress ($\Phi_{\text{grid}} \to 1$), the metric approaches unity (SVI $\to 1$), signaling high eco-debt accumulation where grid stability is sacrificed for sub-optimal computational throughput.
- **State B (Architectural Homeostasis):** When the system achieves full batch saturation ($b(t) \geq b_{\min}$) or executes non-latent workloads during off-peak windows ($\Phi_{\text{grid}} \to 0$), the metric resolves to zero (SVI $\to 0$), indicating the computation is safely buffered by the biophysical capacity of the ecosystem.

## D. *Socio-Technical Regenerative Steering Loops: Metabolic Governance and Architectural Homeostasis*

Replacing linear models with circular systems thinking, the RST reference architecture introduces socio-technical regenerative steering loops governed by the SVI . This governor acts as a cybernetic steering engine, dynamically balancing real-time orchestration (e.g., KV cache allocation and Roofline thresholds) with the macro-level environmental boundaries (PUE/WUE tolerances and CBAM/CSDDD compliance pathways) shown in the outer area of Fig. 1. This homeostatic orchestrator alters execution profiles to minimize thermodynamic entropy when the SVI exceeds its regulatory ceiling. It either bundles fragmented requests into efficient workloads ($b \geq b_{\min}$) or automatically shifts non-latent processing tasks to decentralized regional energy micro-grids.

Over medium-to-long horizons, this governor mitigates the feedback loops and cascading SoS failures. At the core execution layer, scaling context horizons ($L$) causes linear Key-Value cache memory sprawl ($M_{\text{KV}}$). Left unregulated, this fragmentation forces orchestration layers to shard workloads across multi-GPU clusters purely to pool HBM capacity rather than harvest raw compute engines. This fragmentation traps the hardware in a high-entropy, low-concurrency regime ($b \ll b_{\min}$) that draws near-maximum static baseline power while idling on memory-bus transfers. To sustain latency under these conditions, operators are trapped in a cycle of exponential capital expenditure, constructing additional hyperscale facilities that compound regional grid stress ($\Phi_{\text{grid}}$). By embedding these feedback circuits directly into the compute scheduler, the RST reference architecture establishes an operational structural coupling where regulatory mandates and planetary boundaries act as the definitive homeostatic regulators of the system.

## E. *Regenerative Accounting of Environmental and Social Externalities: Achieving Operational Closure*

By mapping hardware execution through closed metabolic circuits, the RST reference architecture translates abstract sustainability concepts into quantifiable engineering vectors, with non-exhaustive examples below:

- **Memory Wall Carbon ($b_{\min}$):** Minimizing data transport directly minimizes localized thermal dissipation, fulfilling SSbD principles at the silicon layer.
- **Batch Size Economics:** Operating in low-batch regimes ($b < b_{\min}$) divides the baseline energy cost across tokens, rendering inference carbon-intensive.
- **Lifecycle Amortization:** The volume of inference tokens generated over a hardware lifespan must offset embodied carbon liabilities; rapid model deprecation turns this cycle into an ecological liability.
- **Thermal Constraints ($M_{\text{KV}}$):** Extreme bandwidth penalties force ultra-dense multi-GPU packing, penalizing Water Usage Effectiveness (WUE) and generating municipal resource depletion.

By mapping relevant parameters as above, the RST reference architecture translates abstract sustainability concepts into four quantifiable engineering and metrology vectors to support DPPs [5], [6], [7] and automated industrial engineering and auditing.

To sum up, the RST reference architecture offers a more scalable and controllable model to manage physical limits of thermodynamics and material scarcity.

# V. Discussion

## A. Answering the Research Questions

The integrated Regenerative Socio-Technical (RST) framework bridges the gap between micro-level hardware telemetry and macro-systemic environmental governance by accounting for compute costs within SPaC loops. The results reveal that local optimization for client latency enforces an aggregate energy penalty across the macro-grid. When workloads operate beneath the roofline balance point, the system experiences severe arithmetic underutilization, drawing near-peak baseline thermal power to sustain continuous HBM weight fetches across memory buses. This operational reality exposes the extractive political economy of Nvidia's linear five-layer cake model, which organizes the computing stack as a unidirectional hierarchy that externalizes massive power and infrastructure burdens onto regional public utilities and volatile energy geographies.

Furthermore, multi-dimensional scaling of context windows forces infrastructure layers into memory-driven hardware fragmentation. Workloads are sharded across multi-GPU clusters purely to pool physical memory capacity, trapping the hardware in a low-concurrency regime that continuously expends near-peak baseline power while idling. To maintain throughput under these constraints, hyperscale data centers invest massive CapEx to build redundant facilities. Because each new facility inherits identical idling penalties, material waste, GHG footprints, and localized grid stress compound exponentially.

To break this cycle, the proposed reference architecture shifts the paradigm from an extractive "electrons-to-tokens" model toward an "atoms-to-values" closed loop. Rather than treating waste and thermal entropy as ignored externalities, they are internalized as institutional metabolic liabilities. Parameterizing waste at the SoS level transitions industrial practice from qualitative CSR reporting toward quantitative, real-time auditing. Utilizing DPPs and RegTech to track the GPU lifecycle and capture cooling telemetry allows operators to route water and heat waste directly back into the circularity of local ecosystems.

## B. Match and Contribution

This study contributes to the DCE literature by providing the first SoS Regenerative Reference Architecture within SPaC relationship loops, governed by quantified boundary conditions. This contribution matches three objectives:

1. **Emerging Technology Management:** Traditional frameworks evaluate Generative AI infrastructure as a weightless asset, masking its biophysical footprint. This work reframes accelerator arrays as an interconnected SoS with thermodynamic liabilities. This "system-to-planet" vantage point equips technology managers with architectural tools to navigate macro-grid volatility induced by frontier LLMs.

2. **Practical Frameworks:** To bridge abstract sustainability goals and ground-truth engineering, this study introduces an actionable framework rooted in Regenerative Accounting, SPaC, and STS theory. Controlled by the SVI, it operationalizes DPPs and RegTech to automate real-time resource routing, shifting industrial practice toward quantitative, enforceable auditing.

3. **Value Creation and Alignment:** The RST model redefines industrial value creation away from brute-force scaling toward circular efficiency, collective social learning, and regional resilience. Centered on a "Values/Needs" core, it permits socio-ethical specifications—such as the *Magnifica Humanitas* encyclical [15]—to protect human dignity. Boundary conditions can then involve rare-earth extraction networks and data annotation supply chains based on community consensus.

To transition these insights to institutional adoption, engineering managers can utilize the actionable compliance checklist below:

- **Grid and Energy Community:** Respect regional grid capacity and local energy community needs.
- **Thermodynamic Loop:** Structurally explore waste-to-value loops, using IRDS heat recovery KPIs.
- **Lifecycle Circularity:** Use DPPs to track GPU lifecycles, component refurbishment, and e-waste processing against SSbD and CSDDD guidelines.
- **Asset Amortization:** Amortize hardware pre-training and operational footprints over realistic, extended lifecycles.
- **Supply Chain Ethics:** Actively audit labor conditions, including the rare-earth extraction and data annotation.

# VI. Conclusion: Roadmap for Regenerative AI

This section concludes with a technical roadmap establishing a DCE that is powered by Regenerative AI that remains tethered to planetary realities.

## A. Limitations

While this study introduces a holistic RST reference architecture for token-driven AI, it faces three limitations mapped at different Technology Readiness Levels (TRLs):

- **Conceptual Stage (TRL 1–2):** The framework has yet to be adopted **by** official IEEE IRDS International Focus Teams (IFTs) for formal industry roadmapping.
- **Lack of Micro-Telemetry:** While macroscopically mapping material and thermodynamic flows (TRL 9 hardware constraints) onto SPaC loops, the study lacks component-level requirement guidelines or physical hardware telemetry specifications.
- **Validation Needs (TRL 2–3)**: The proposed RA protocols and SVI require extensive peer review and large-scale pilot implementations of DPPs across diversified global supply chains (GVC) to verify distributed data integrity for real-time auditing.

## B. Concluding Remarks: Computes within Boundaries

This study contributes a multi-disciplinary critique of the energy-intensive trajectories of the 2025–2026 token-based economy, shifting from linear "Device-to-System" scaling to a circular "System-to-Planet" relationship. Our main finding shows that without a "Circular Governor" enabled by RA and DPPs, the AI industry faces an imminent "Hard Landing" against planetary limits. Isolating engineering efficiency from metabolic consumption accelerates resource depletion, validating Jevons' Paradox.

This misalignment is currently visible in Taiwan: Taipower has suspended new power applications for data centers exceeding 5 MW north of Taoyuan to prevent localized grid failure [14], This occurs at the exact juncture where Nvidia has established its "Constellation" regional headquarters in the Beitou-Shilin Technology Park, triggering explicit leadership warnings regarding the island's binding energy limits [13].

The proposed RST reference architecture provides the IEEE community with an initial blueprint needed to internalize ecological debt as a core technical requirement.

### C. Future Work/Perspectives

Based on our vision for the Safe-and-Sustainable-by-Design (SSbD) Chips Value Chain Services (SCVCS) Research Initiative [55], future research will investigate how to transform chips and AI supply chains into environmentally accountable, circular networks by using AI orchestrators, trustworthy autonomous agents, and digital product passports to balance high manufacturing yields with strict sustainability and regulatory compliance. To achieve this, future work must bridge the gap between reference models and technical objectives across the IEEE roadmap ecosystem:

- **Power and Energy Socio-Technical Alignment**: The 2024 *IEEE Power and Energy Technology Assessment* and *the 2023 Control for Societal-scale Challenges* should set strategic directions on energy communities and hyperscale facilities, interacting with the IEEE Heterogeneous Integration Roadmap (HIR), specifically Chapters 2 (High-Performance Computing), 8 (Advanced Manufacturing), and 10 (Integrated Power Electronics).
- **Materials and Chip Supply Chain Alignment:** Annual IEEE IRDS activities should apply an SoS approach to monitor material-chips-hyperscale value chains, deploying DPPs to track material and chip lifecycles, and considering the Regenerative Accounting framework.
- **Energy Community Expansion and Inter-Agency Action:** The IEEE International Roadmap Committee (IRC) should charter a "Socio-Technical Governance" domain to analyze infrastructure through the Planetary Boundary Resilience lens. This system thinking should shape agendas like the United for Smart Sustainable Cities (U4SSC) into a new coalition: United for Smart Sustainable Intelligence (U4SSI) to help leaders and investors align value propositions that foster resilient, circular systems.

Ultimately, deep technologies demand deep systems thinking anchored in reality. Transforming generative AI into a truly regenerative asset requires standardizing actionable technical frameworks for real-time, enforceable regenerative accountability of both energy and compute resources.

## ACKNOWLEDGMENT

The authors gratefully acknowledge the constructive feedback of the double-blind peer reviewers whose scrutiny strengthened this work; all remaining errors and interpretations are solely the authors' own responsibility.